\documentclass[aps,twocolumn]{revtex4}
\usepackage{graphicx}

\begin{document}

\title{Pressure Tuned Enhancement of Superconductivity and Change of Ground State Properties in
LaO$_{0.5}$F$_{0.5}$BiSe$_2$ Single Crystals}

\author{Jianzhong Liu, Sheng Li, Yufeng Li, Xiyu Zhu$^{*}$, Hai-Hu Wen}\email{zhuxiyu@nju.edu.cn, hhwen@nju.edu.cn}

\affiliation{Center for Superconducting Physics and Materials,
National Laboratory of Solid State Microstructures and Department
of Physics, National Center of Microstructures and Quantum
Manipulation, Nanjing University, Nanjing 210093, China}

\date{\today}

\begin{abstract}
By using a hydrostatic pressure, we have successfully tuned the
ground state and superconductivity in LaO$_{0.5}$F$_{0.5}$BiSe$_2$
single crystals. It is found that, with the increase of pressure, the original superconducting
phase with $T_c$ $\sim$ 3.5 K can be tuned to a state with lower $T_c$, and then a new
superconducting phase with $T_c$ $\sim$ 6.5 K emerges. Accompanied by this
crossover, the ground state is switched from a semiconducting
state to a metallic one. Accordingly, the normal state resistivity also shows a
nonmonotonic change with the external pressure. Furthermore, by
applying a magnetic field, the new superconducting state under pressure with
 $T_c$ $\sim$ 6.5 K is suppressed, and the normal state reveals a weak
semiconducting feature again. These results illustrate a
non-trivial relationship between the normal state property and
superconductivity in this newly discovered superconducting system.


\end{abstract} \maketitle

According to the theory of Bardeen-Cooper-Schrieffer (BCS),
superconductivity is achieved by the quantum condensation of
Cooper pairs which are formed by the electrons with opposite
momentum near the Fermi surface. The ground state when
superconductivity is removed is thus naturally believed to be metallic. In some unconventional superconductors, such as cuprate,
iron pnictide/phosphide, heavy fermion and organic
superconductors, this may not be true. Recently, the BiS$_2$-based
superconductors whose structures are similar to the
cuprates\cite{Cuprates} and iron pnictides\cite{Iron-based}, have
been discovered and formed a new superconducting (SC) family. Many new SC compounds
with the BiS$_2$ layer have been found, including
Bi$_4$O$_4$S$_3$\cite{BiOS1,Awana BiOS,Li BiOS},
REO$_{1-x}$F$_x$BiS$_2$ (RE=La, Nd, Ce, Pr and
Yb)\cite{LaOBiS,NdOBiS,X.J CeOBiS,PrOBiS,Maple LnOBiS},
Sr$_{1-x}$La$_x$FBiS$_2$\cite{SrLaFBiS} and
La$_{1-x}$M$_x$OBiS$_2$ (M=Ti, Zr, Hf and Th)\cite{LaMOBiS}, etc.. Among
these compounds, the high pressure synthesized
LaO$_{0.5}$F$_{0.5}$BiS$_2$ was reported to have a maximum $T_c$
$\sim$ 10.6 K\cite{LaOBiS}. The basic band structure obtained by
first principle calculations indicates the presence of strong
Fermi surface nesting at the wave vector ($\pi$,$\pi$)\cite{BiS
theory,WanXianGang theory,Yildirim theory} when the doping is
close to x=0.5 in, for example, LaO$_{1-x}$F$_x$BiS$_2$. Quite
often the superconductivity is accompanied by a normal state with
a clear semiconducting behavior with unknown
reasons\cite{LaOBiS,NdOBiS,X.J CeOBiS,PrOBiS,Maple
LnOBiS,SrLaFBiS,LaMOBiS}. In addition, possible triplet paring and
weak topological superconductivity were suggested based on
renormalization-group numerical calculation\cite{triplet}, but
this mechanism is still much debated. Moreover, the experiments on
the NdO$_{0.5}$F$_{0.5}$BiS$_2$ single crystals also reveal
interesting discoveries concerning the SC mechanisms in this new
system\cite{NdOBiS Single crystal,Liu single crystal,DL Feng
ARPES,Ding Hong ARPES,HQ yuan}.

Through adjusting the lattice parameters and intimately the
electronic band structure, high pressure has served as a very
effective method, which can tune both the SC and normal state of
superconductors. In this newly found BiS$_2$ family, high pressure
has been recognized as an important tool to enhance both the
superconductivity volume and transition temperatures except for
Bi$_4$O$_4$S$_3$\cite{P BiOS LaOBiS,Maple P La/CeOBiS,Maple P
Pr/NdOBiS,Awana P SrLaFBiS,Awana P CeOBiS,up to 18Gpa,Awana P
SrReFBiS2}. In particular, the SC transition temperature of
REO$_{1-x}$F$_x$BiS$_2$ (RE=La, Ce, Nd, Pr)\cite{P BiOS
LaOBiS,Maple P La/CeOBiS,Maple P Pr/NdOBiS} and
Sr$_{1-x}$RE$_x$FBiS$_2$ (RE = La, Ce, Nd, Pr, Sm)\cite{Awana P
SrLaFBiS,Awana P SrReFBiS2} systems was enhanced tremendously by
applying the hydrostatic pressure. Taking
LaO$_{0.5}$F$_{0.5}$BiS$_2$ as an example, the $T_c$ of the sample
can be increased from about 2 K under ambient pressure to $\sim$
10 K under 2 GPa\cite{P BiOS LaOBiS}. And in the
Sr$_{1-x}$RE$_x$FBiS$_2$ (R= Ce, Nd, Pr, Sm) system, the non-SC
sample at ambient pressure can also be tuned to become a SC one
with $T_c$ $\sim$ 10 K under a pressure of 2.5 GPa\cite{Awana P
SrReFBiS2}. To understand the role of high pressure,
X-ray diffraction measurements under pressures have been performed on
LaO$_{0.5}$F$_{0.5}$BiS$_2$ system and suggest a structural phase
transition from a tetragonal phase ($P4/nmm$) to a monoclinic
phase ($P21/m$) under pressures\cite{up to 18Gpa}. Very recently,
a new superconductor LaO$_{0.5}$F$_{0.5}$BiSe$_2$ with the same
structure as the LaO$_{0.5}$F$_{0.5}$BiS$_2$ was discovered
with $T_c$ $\sim$ 3.5 K\cite{LaOBiSe,LaOBiSe single crystal
1,LaOBiSe single crystal 2}. It was reported that the electronic
structure and Fermi surface in these two compounds are quite
similar\cite{BiSe2 theory}. Since the system now is selenium
based, it is highly desired to do investigations on BiSe$_2$-based
materials, better in form of single crystals. Furthermore it is
curious to know how the high pressure influences the
superconductivity and the ground state property in the
BiSe$_2$-based superconductors.

Here, we report the successful growth of the
LaO$_{0.5}$F$_{0.5}$BiSe$_2$ single crystals, and a systematic
high-pressure study on two single crystals (hereafter named as Sample-1 and Sample-2). By increasing pressure, the
ground state is switched from a semiconducting state to a metallic
one, simultaneously the original SC $T_c$ $\sim$ 3.5 K (at ambient
pressure) initially drops down to about 2 K and finally increases
with pressure. As the pressure reaches about 1.2$\pm$0.2 GPa, a new SC phase
with higher $T_c$ appears. At about 2.17 GPa, the $T_c$ of the new SC
phase reaches about 6.5 K. Accompanied with the change of SC
transition temperatures, the normal state resistivity ($\rho_n$) decreases first and then increases with pressure. This non-monotonic pressure dependence of $T_c$ and the normal state resistivity in the present BiSe$_2$-based system are very different from the BiS$_2$-based family. Furthermore, the SC phase with higher T$_c$ can be suppressed by applying a
magnetic field, and a weak semiconducting feature in the normal
state emerges again when superconductivity is suppressed. All
these results show the competing feature between superconductivity
and the underlying ground state associated with the semiconducting behavior.

\begin{figure}
\includegraphics[width=8.5cm]{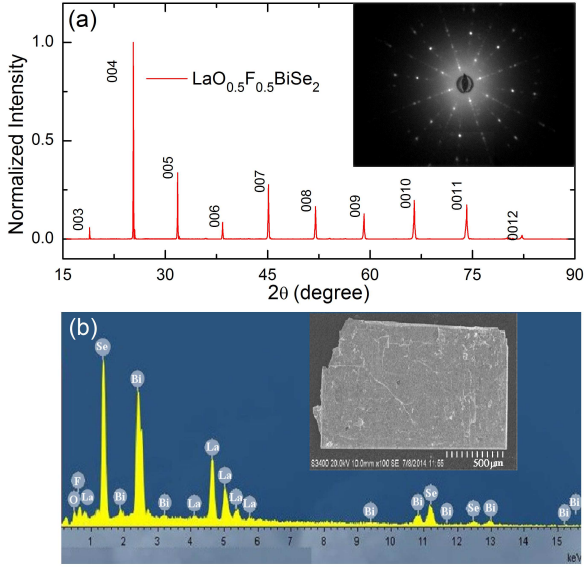}
\caption{(Color online) (a) X-ray diffraction pattern for a
LaO$_{0.5}$F$_{0.5}$BiSe$_2$ single crystal. The inset shows the back
Laue X-ray diffraction pattern of a LaO$_{0.5}$F$_{0.5}$BiSe$_2$
single crystal. (b) Energy Dispersive X-ray microanalysis spectrum
taken on a LaO$_{0.5}$F$_{0.5}$BiSe$_2$ single crystal. The inset
shows the SEM photograph of the crystal with typical dimensions of
about $1.4\times0.7\times0.04$ mm$^3$.} \label{fig1}
\end{figure}

The LaO$_{0.5}$F$_{0.5}$BiSe$_2$ single crystals were grown by using flux
method\cite{Liu single crystal}. Powders of La$_2$O$_3$, LaF$_3$,
Bi$_2$Se$_3$, Se and La scraps (all 99.9\% purity) were mixed in
stoichiometry as the formula of LaO$_{0.5}$F$_{0.5}$BiSe$_2$. The
mixed powder was grounded together with CsCl/KCl powder (molar
ratio CsCl : KCl : LaO$_{0.5}$F$_{0.5}$BiSe$_2$ = 12 : 8 : 1) and
sealed in an evacuated quartz tube. Then it was heated up to
800$^\circ$C for 50 hours followed by cooling down at a rate of
3$^\circ$C/hour to 600$^\circ$C. Single crystals with lateral
sizes of about 1 mm were obtained by washing with water. X-ray
diffraction (XRD) measurements were performed on a Bruker D8
Advanced diffractometer with the Cu-K$_\alpha$ radiation. DC magnetization
measurements were carried out with a SQUID-VSM-7T (Quantum
Design).

\begin{figure}
\includegraphics[width=8.2cm]{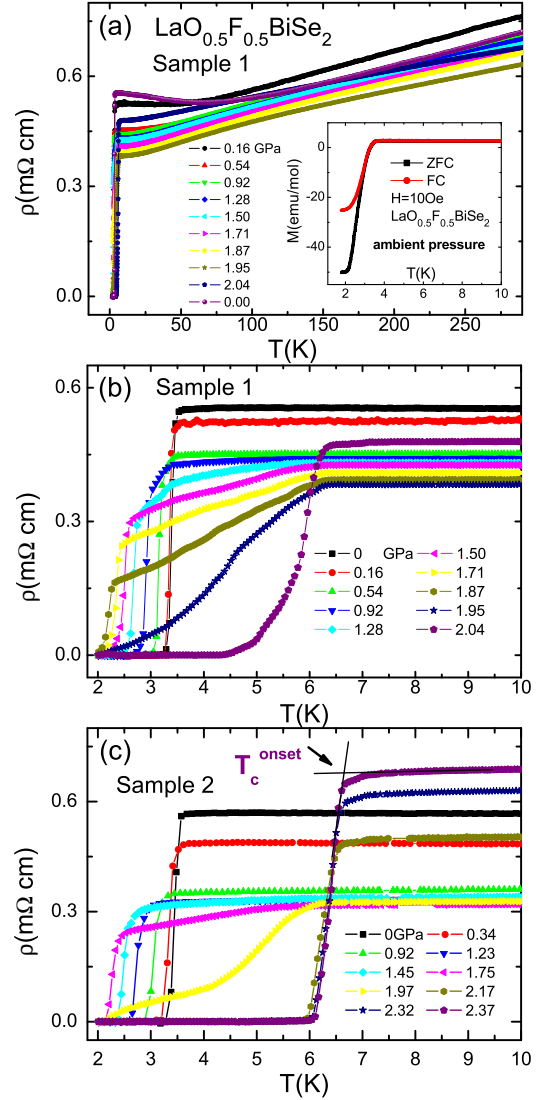}
\caption{(Color online) (a) Temperature dependence of resistivity
for Sample-1 at various pressures in the temperature range 2 K to
300 K. The inset shows the magnetic susceptibility of a
LaO$_{0.5}$F$_{0.5}$BiSe$_2$ single crystal in an applied field of
10 Oe ($\parallel$c-axis) under the ambient pressure. Both the
magnetic susceptibility measured in zero-field-cooled (ZFC) and
field-cooled (FC) modes are shown. (b) and (c) Enlarged views of
the resistive transition in the temperature range 2 K to 10 K at
various pressures for Sample-1 and Sample-2, respectively. The
superconducting transitions are rather sharp at ambient and high
pressures.}\label{fig2}
\end{figure}

Measurements of resistivity under pressure were performed up
to $\sim$ 2.3GPa on PPMS-16T (Quantum Design) by using HPC-33
Piston type pressure cell with the Quantum Design DC resistivity
and AC transport options. The LaO$_{0.5}$F$_{0.5}$BiSe$_2$ single
crystal with the standard four-probe method was immersed in
pressure transmitting medium (Daphene 7373) in a Teflon cap.
Hydrostatic pressures were generated by a BeCu/NiCrAl clamped
piston-cylinder cell. The pressure upon the sample was determined
by measuring the pressure-dependent $T_c$ of a Sn sample with high purity.

In Fig.~\ref{fig1}(a) we present the X-ray diffraction (XRD)
pattern for the LaO$_{0.5}$F$_{0.5}$BiSe$_2$ single crystal. It's
clear that only ($00l$) reflections can be observed yielding a
$c$-axis lattice constant $c=14.05\pm0.03\AA$. The inset of
Fig.~\ref{fig1}(a) shows the Laue diffraction pattern of the
LaO$_{0.5}$F$_{0.5}$BiSe$_2$ single crystal. Bright and symmetric
spots can be clearly observed, indicating a good crystallinity.
Energy dispersive X-ray spectrum (EDS) measurements were performed
at an accelerating voltage of 20kV and working distance of 10
millimeters by a scanning electron microscope (Hitachi Co.,Ltd.).
One set of the EDS result on LaO$_{0.5}$F$_{0.5}$BiSe$_2$ single
crystal is shown in Fig.~\ref{fig1}(b), and the composition of the
single crystal can be roughly expressed as
LaO$_y$F$_{0.48}$Bi$_{0.95}$Se$_{1.89}$. The atomic ratio is close
to the nominal composition except for oxygen which can not be obtained
accurately by the EDS measurement.

The temperature dependence of resistivity for the
LaO$_{0.5}$F$_{0.5}$BiSe$_2$ single crystal (Sample-1) at various pressures
with temperature ranging from 2 K to 300 K is illustrated in
Fig.~\ref{fig2}(a). The inset of Fig.~\ref{fig2}(a) shows the
temperature dependent magnetic susceptibility at ambient pressure
under a magnetic field of 10 Oe, and a sharp SC transition is
observed at about 3.5 K. An estimate on the Meissner screening
volume through the magnetic susceptibility measured in the
zero-field-cooled (ZFC) mode reveals a high superconducting volume. For Sample-1, we were not managed to measure the sample at a pressure higher than 2.04 GPa. As shown in Fig.~\ref{fig2}(a), at ambient pressure the
normal state resistivity shows a semiconducting behavior. This
semiconducting behavior can be suppressed under a small pressure
and turns to be a metallic one at about 0.54 GPa. With further increase of pressure, the metallic behavior maintains until the
maximum pressure. This semiconducting to metallic
transition with pressure has been noticed in
Sr$_{1-x}$RE$_x$FBiS$_2$ (R= La, Ce, Nd, Pr, Sm)
systems\cite{Awana P SrLaFBiS,Awana P SrReFBiS2}. In the case of
Sr$_{0.5}$La$_{0.5}$FBiS$_2$ polycrystalline sample, the
semiconductor-metal transition was considered as coming from the
change of F-Sr/La-F bond angle along with inter-atomic
distances\cite{Awana P SrLaFBiS}. Interestingly, the
semiconductor-metal transition under pressure for
CeO$_{0.5}$F$_{0.5}$BiS$_2$ system has been proposed according to
the first-principle calculations\cite{CeOBiS transition from
theroy}, but the transition was not observed in
previous reports of experiment\cite{Maple P La/CeOBiS,Awana P CeOBiS}. In particular,
for LaO$_{0.5}$F$_{0.5}$BiS$_2$ polycrystalline samples, the
normal state resistivity decreases monotonically with increasing
pressure, but it still exhibits semiconducting behavior under a
very high pressure (18 GPa)\cite{up to 18Gpa}.

In Fig.~\ref{fig2}(b) and (c), we present enlarged views of SC transitions at low temperatures under
various pressures for Sample-1 and Sample-2, respectively. Both samples exhibit very similar behavior. As one can see, the variation of both the SC transition temperature and normal state resistivity upon the external pressure are non-monotonic. The original
$T_c$ $\sim$ 3.5 K (at ambient pressure) gradually drops down with
increasing pressure and becomes below 2 K at about 1.95 GPa. At
the same time, a high $T_c$ phase gradually emerges starting from
about 1.2$\pm$0.2 GPa and enhances continuously with increasing pressure. It
seems that the high T$_c$ phase with T$_c$ = 6.5K coexists with the low
$T_c$ phase in the range from 1.2$\pm$0.2 GPa to about 1.95 GPa. With further
increase of pressure, zero resistance corresponding to the high $T_c$
phase appears above 2 K and the SC transition becomes sharper at higher
pressures. A similar behavior under pressure
has been observed in some strongly correlated electronic systems,
such as heavy fermion\cite{CeCu2Si2 P Yuan}, organic
systems\cite{organic P} and iron chalcogenides\cite{KFe2Se2 P}. In
previous high pressure studies on BiS$_2$-based superconductors,
the T$_c$ monotonically increases with the pressure without
showing the coexistence of two transient phases. This indicates
the distinction between our present BiSe$_2$-based superconductors
and the earlier studied BiS$_2$-based systems. For Sample-1 we were not managed to measure the resistivity beyond 2.04 GPa. Two samples
are from the same batch. One can see that the resistive transitions  below 2.04 GPa are quite
similar to each other.

It is worth noting that the normal state resistivity presents
a non-monotonic dependence on applied pressure. As shown in Fig. 2(b) and (c),
the normal state resistivity just above the SC
transition temperature gradually decreases with increasing
pressure till about 1.95 GPa. Surprisingly, above the threshold pressure, the normal
state resistivity begins to increase remarkably with increasing
pressure. It is clear that this qualitative behavior is
closely related to the pressure-dependent $T_c$, as we addressed below.

\begin{figure}
\includegraphics[width=8.5cm]{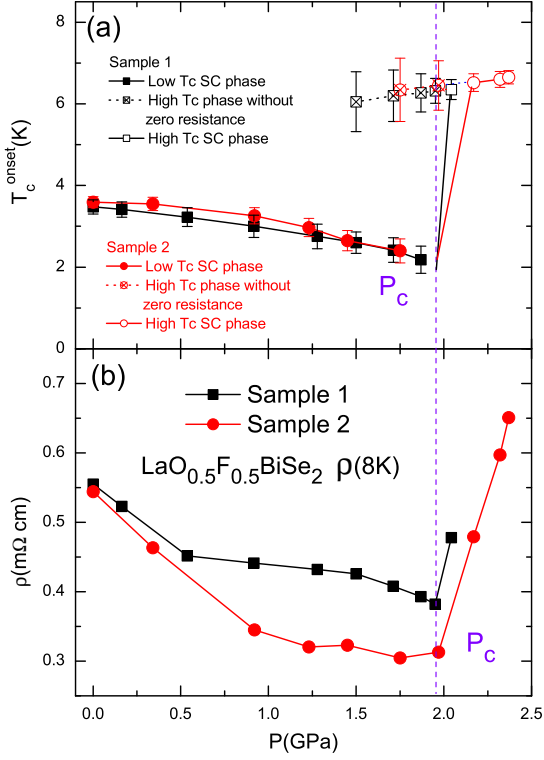}
\caption{(Color online)(a) Phase diagram of $T_c^{onset}$ versus
pressure for the  two LaO$_{0.5}$F$_{0.5}$BiSe$_2$ single crystals
investigated here. The dark and red symbols represent the
$T_c^{onset}$ of Sample-1 and Sample-2, respectively. The filled
symbols stand for the low $T_c$ phase, the open and crossed
symbols stand for the high $T_c$ phase with and without zero
resistance, respectively. (b) Resistivity at 8 K in the normal
state at various pressures for Sample-1 (filled squares) and
Sample-2 (filled circles) .} \label{fig3}
\end{figure}

Fig.~\ref{fig3}(a) and ~\ref{fig3}(b) present the phase
diagram of $T_c^{onset}$ versus pressure and pressure-dependent
resistivity (8K), respectively. Here, the pressure for the absence
of the second transition (about 1.95GPa) is defined as the critical
one ($P_c$). Fig.~\ref{fig3}(a) and ~\ref{fig3}(b) clearly
reveal two distinct SC phases: the low $T_c$ SC phase below $P_c$ and
the high $T_c$ SC phase above $P_c$. In the low $T_c$ SC phase region, both $T_c$ and the normal state resistivity are suppressed
with increasing pressure. On the contrary, in the high $T_c$ SC
phase, $T_c$ is slightly enhanced and the normal state resistivity
increases remarkably with raising pressure. In
LaO$_{0.5}$F$_{0.5}$BiS$_2$ polycrystalline samples, a structural
phase transition from a tetragonal phase ($P4/nmm$) to a
monoclinic phase ($P21/m$) has been suggested by high-pressure
X-ray diffraction measurements. And a high $T_c$ value of 10.7K in
the high-pressure regime appears in the monoclinic
structure\cite{up to 18Gpa}. Therefore, considering the very weak change of the transition temperature of the high $T_c$ phase, and taking a comparison between LaO$_{0.5}$F$_{0.5}$BiS$_2$ and LaO$_{0.5}$F$_{0.5}$BiSe$_2$, we believe that
there are two distinct SC phases in our
LaO$_{0.5}$F$_{0.5}$BiSe$_2$ single crystals in the intermediate pressure region (1.2$\pm$0.2 to $\sim$ 1.95 GPa). At a high
pressure, all the phase becomes superconductive with T$_c$ $\sim$
6.5 K. The transition from the low $T_c$ phase to the high $T_c$ one could be induced by the structural transition, which needs to be further checked.

\begin{figure}
\includegraphics[width=8.5cm]{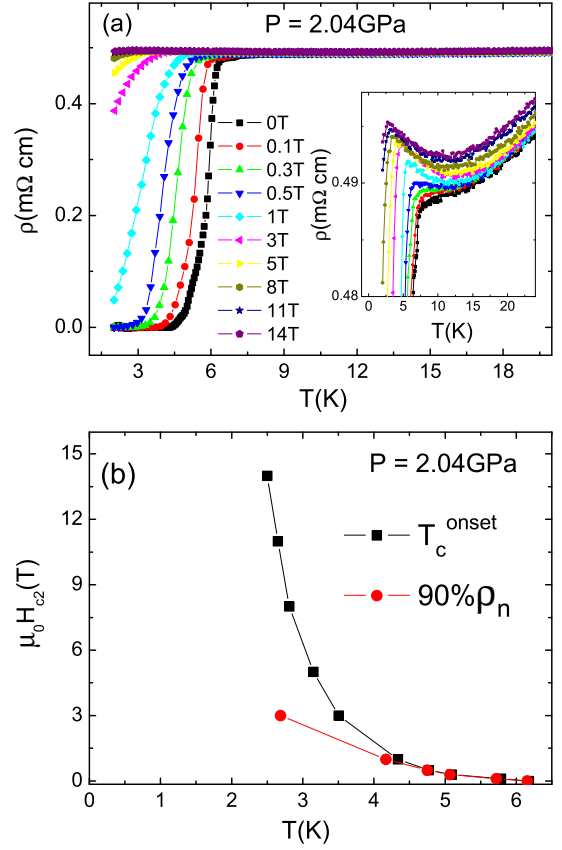}
\caption{(Color online) (a) Temperature dependence of resistivity
for Sample-1 under a pressure of 2.04 GPa at various magnetic
fields. The inset shows the enlarged view of a weak semiconducting
feature in the normal state under high magnetic fields. (b) Upper
critical field determined by $T_c^{onset}$ and 90\% normal state
resistivity $\rho_{n}$.} \label{fig4}
\end{figure}

  In Fig.~\ref{fig4}(a), we present the temperature dependent resistivity under
magnetic field up to 14 T at 2.04 GPa ($T_c$ $\sim$ 6.3K). The
upper critical field $H_{c2}$ versus $T_c$ is displayed in
Fig.~\ref{fig4}(b). We use different criterions of 90\%$\rho_{n}$
and $T_c^{onset}$ (determined using the crossing point shown in Fig.~\ref{fig2}(c)) to determine the $H_{c2}$. The upper critical
field at zero temperature can be estimated by using the
Werthamer-Helfand-Hohenberg (WHH) formula\cite{WHH formula}
${H_{c2}=-0.69T_{c}[dH_{c2}/dT]_{T_c}}$, and the estimated
$H_{c2}(0)$ is about 35 T for $T_c^{onset}$. The inset of Fig.~\ref{fig4}(a) shows the enlarged
view of superconducting transitions as in the main panel. As
shown in the inset, the SC is very robust and keeps presence above
2 K when the field is up to 14 T. That could be induced by the fact that the applied
field was approximately parallel to $ab$ plane of the single
crystal in the pressure cell during the measurement, and a large anisotropy has been discovered in LaO$_{0.5}$F$_{0.5}$BiSe$_2$ single crystals\cite{LaOBiSe
single crystal 1}. An interesting phenomenon is that a weak
semiconducting behavior re-emerges when the superconductivity is
suppressed under a high magnetic field. A similar behavior was
observed in NdO$_{0.5}$F$_{0.5}$BiS$_2$ single crystals\cite{Liu
single crystal}. This phenomenon may be related to the
semiconducting behavior of the sample at an ambient pressure,
although it seems that the low T$_c$ phase does not show up here.
The semiconducting ground states for either the low T$_c$ phase at
an ambient pressure, or the one with high T$_c$ superconductivity
under a high pressure but suppressed with a high magnetic field,
may be caused by the same reason, both point to the competition of
superconductivity with a tendency which underlines the
semiconducting behavior.

In summary, we have successfully tuned the ground state and
superconductivity in LaO$_{0.5}$F$_{0.5}$BiSe$_2$ single crystals
through a hydrostatic pressure. The ground state is switched from
a semiconducting state to a metallic one with increasing pressure.
Moreover, the original SC phase with $T_c$ $\sim$ 3.5 K can be
tuned to a new SC state with $T_c$ $\sim$ 6.5 K. In the low $T_c$ SC phase region, both $T_c$
and the normal state resistivity are suppressed with increasing
pressure. On the contrary, in the high $T_c$ SC phase,
superconductivity is enhanced and the normal state resistivity
increases remarkably with increasing pressure. Moreover, a weak
semiconducting behavior re-emerges when the superconductivity
under a high pressure is suppressed under magnetic field. These
results illustrate a non-trivial relationship between the normal
state property and superconductivity. Further theoretical and detailed structure
investigations are highly desired to clarify the new high $T_c$ SC phase under a high pressure.

\section*{ACKNOWLEDGMENTS}
We appreciate the kind help and discussions
with Xiaojia Chen. We thank Xiang Ma and Dong Sheng for the assistance in SEM/EDS
measurements. This work was supported by NSF of China, the
Ministry of Science and Technology of China (973 projects:
2011CBA00102, 2012CB821403, 2010CB923002) and PAPD.

\end{document}